\pdfoutput=1
\newif\ifarxiv
\arxivtrue

\DocumentMetadata{}
\ifarxiv
\documentclass[acmsmall,screen,nonacm,nonacm,dvipsnames,svgnames]{acmart}
\else
\documentclass[acmsmall,screen,review,nonacm,dvipsnames,svgnames]{acmart}
\fi

\ifarxiv
\setcopyright{cc}
\setcctype[4.0]{by}
\else
\renewcommand\footnotetextcopyrightpermission[1]{}
\fi

\usepackage{xcolor,soul}
\usepackage[noline,noend]{algorithm2e}
\usepackage{listings}
\usepackage[shortlabels]{enumitem}

\usepackage[capitalize,nameinlink]{cleveref}
\usepackage{caption, subcaption, stackengine}
\usepackage[utf8]{inputenc}
\usepackage{mathtools}
\usepackage{qtree}
\usepackage{tikz-cd}
\usepackage{wrapfig}
\usepackage{float}
\usepackage{forest}

\definecolor{GrayCodeBlock}{RGB}{241,241,241}
\definecolor{BlackText}{RGB}{110,107,94}
\definecolor{BlueTypename}{RGB}{17,86,182}
\definecolor{RedTypename}{RGB}{182,86,17}
\definecolor{GreenString}{RGB}{96,172,57}
\definecolor{PurpleKeyword}{RGB}{184,84,212}
\definecolor{GrayComment}{RGB}{170,170,170}
\definecolor{GrayNumber}{RGB}{200,200,200}
\definecolor{GoldDocumentation}{RGB}{180,165,45}
\ifarxiv
\else
\fi

\AtBeginDocument{%
  \providecommand\BibTeX{{%
    \normalfont B\kern-0.5em{\scshape i\kern-0.25em b}\kern-0.8em\TeX}}}

\begin{document}

\title{Towards Relational Contextual Equality Saturation}

\author{Tyler Hou}
\orcid{0009-0002-2234-2549}
\email{tylerhou@berkeley.edu}
\affiliation{%
  \institution{UC Berkeley}
  \city{Berkeley}
  \state{CA}
  \country{USA}
}

\author{Shadaj Laddad}
\orcid{0000-0002-6658-6548}
\email{shadaj@cs.berkeley.edu}
\affiliation{%
  \institution{UC Berkeley}
  \city{Berkeley}
  \state{CA}
  \country{USA}
}

\author{Joseph M. Hellerstein}
\orcid{0000-0002-7712-4306}
\email{hellerstein@cs.berkeley.edu}
\affiliation{%
  \institution{UC Berkeley}
  \city{Berkeley}
  \state{CA}
  \country{USA}
}

\newcommand{\egglog}{{\small \tt egglog}}
\newcommand{\egg}{{\small \tt egg}}
\newcommand{\Rainbow}{Rainbow}

\newcommand{\join}{\sqcup{}}
\newcommand{\meet}{\sqcap{}}
\newcommand*{\defeq}{\stackrel{\text{\tiny def}}{=}}
\let\emptyset\varnothing

\newcommand{\ID}{\textsc{id}}
\newcommand{\eq}{\textit{eq}}
\newcommand{\ctx}{\textit{ctx}}
\newcommand{\roott}{\textit{root}}
\DeclarePairedDelimiter\abs{\lvert}{\rvert}

\begin{abstract}
  Equality saturation is a powerful technique for program optimization.
  Contextual equality saturation extends this to support rewrite rules that are
  conditioned on where a term appears in an expression. Existing work has brought
  contextual reasoning to egg; in this paper, we share our ongoing work to extend
  this to relational equality saturation in egglog. We summarize the existing
  approaches to contextual equality saturation, outline its main applications,
  and identify key challenges in combining this approach with relational models.
\end{abstract}


\maketitle

\section{Introduction}

In recent years there as been a surge of interest in using equality saturation
to optimize programs \cite{eqsat}. Equality saturation is an optimization
technique which repeatedly applies rewrite rules to an program. These rewrite
rules add equalities between (sub)terms of the expression to a data structure
called an e-graph. The e-graph represents a possibly-infinite set programs
equivalent to the original program. Finally, an optimized program can be
extracted from the e-graph.

However, existing equality saturation tools like \egg{} \cite{egg} and
\egglog{} \cite{egglog} do not natively support {\it contextual equalities}.
Contextual equalities are equalities between terms that are valid in some
subgraph, but may not be valid in general. For example, consider the ternary $x
  == 2\;?\;(x \times y) : y$. Under the {\tt then} branch of the ternary, the
condition $x == 2$ is satisfied, so an optimizer could replace the
multiplication with a bitshift, producing an optimized expression $x ==
  2\;?\;(y \gg 1) : y$. However, equating the term $x \times y$ with $y \gg 1$ in
general is not valid, because $x \times y$ may appear in some other term where
the condition is not satisfied.

Reasoning under contexts is necessary in many program optimizers to achieve
better performance. For example, by applying context-aware optimizations to
circuit design, an RTL optimization tool reduced circuit area by 41\% and delay
by 33\% \cite{egg-datapath}. In addition, reasoning about contextual properties
like sort order is critical for dataflow optimization and relational query
optimization \cite{stateful-dataflow, cascades}.

\section{Existing Approaches}

There have been various approaches to support contextual reasoning in equality
saturation frameworks. The authors of the RTL optimization tool encoded
contextual reasoning in \egg{} by adding {\tt ASSUME(x, c)} e-nodes to the
e-graph, where {\tt x} is the expression to be optimized, and {\tt c} is a set
of expressions which represent known constraints \cite{egg-datapath}. Auxiliary
rewrite rules ``push down'' {\tt ASSUME} e-nodes into the expression tree,
adding additional constraints when appropriate. Finally, after {\tt ASSUME}
e-nodes have been pushed down sufficiently far, domain-specific rules rewrite
them into more efficient terms using the constraints. However, the main
limitation of this approach is that these ``extra'' {\tt ASSUME} e-nodes
significantly expand the size of the e-graph, harming performance. In their
tool, optimization of a floating point subtactor took 22 minutes, with the
majority of time spent in e-graph expansion \cite{egg-datapath}.

An alternative to adding explicit {\tt ASSUME} e-nodes into the e-graph is to
annotate e-classes with the contexts that they appear in via a top-down
analysis. Then, a contextual rewrite rule can ``copy'' the contextual subgraph,
replacing subterms that can be contextually rewritten with their equivalents
\cite{alex-thesis}. However, this approach can also substantially expand the
e-graph: to avoid false equalities, such a rewrite must make independent copies
of all contextually-rewritten e-classes and their ancestors, up to and
including the ``source'' of the context. When multiple contexts are nested,
this can again lead to a combinatorial explosion of the e-graph size. Finally,
copying the e-graph also makes it more expensive to apply context insensitive
rewrites -- such rewrites may have to match all copied subgraphs since they are
no longer related to the original subgraph.

A promising approach for efficiently supporting contextual equalities are
colored e-graphs \cite{colored-egraphs}. Instead of adding new e-nodes to the
e-graph, colored e-graphs support multiple equivalence relations: a base
equivalence relation represents equalities that apply in every context, and
context-sensitive, colored equivalences are {\it layered} on top of the base
relation. These layered relations can be thought of ``shallow'' copies of the
base relation. This approach has two advantages: first, it saves memory because
context-sensitive rewrites typically add relatively few equivalences on top of
the base equivalence. Second, when additional base equivalences are found, the
context-sensitive equivalence relations can be efficiently updated. However,
one limitation of the colored e-graph work is that it targets the \egg{}
library, which implements non-relational e-matching.

\section{Case studies}

To better understand the space of applications that can benefit from contextual
equality saturation, we explore three case studies from a range of optimization
domains.

\newcommand{\edges}{\text{e}}

\subsection{Relational query optimization}
\label{sec:query-opt}
\begin{figure}[h]
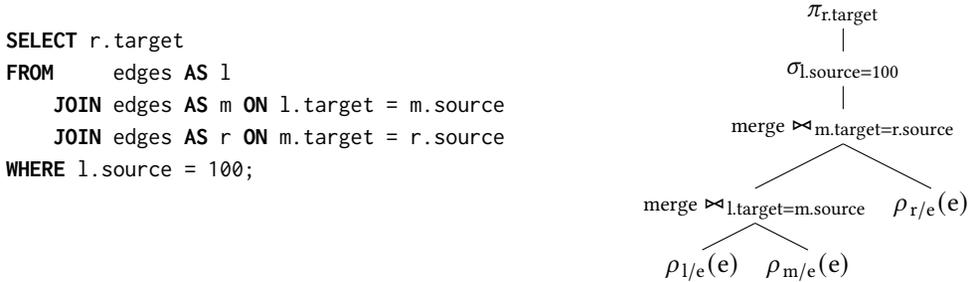

  \begin{subfigure}[t]{0.6\linewidth}
    \begin{lstlisting}[language=sql, basicstyle=\small\ttfamily, basewidth=0.5em, numbers=none]
SELECT r.target
FROM     edges AS l
    JOIN edges AS m ON l.target = m.source
    JOIN edges AS r ON m.target = r.source
WHERE l.source = 100;
        \end{lstlisting}

  \end{subfigure}%
  \begin{subfigure}[t]{0.4\linewidth}
    \Tree[.{$\pi_{\text{r.target}}$}
        [.{$\sigma_{\text{l.source} = 100}$}
            [.{$\text{\smaller merge} \bowtie_{\,\text{m.target}=\text{r.source}}$}
                [.{$\text{\smaller merge} \bowtie_{\,\text{l.target}=\text{m.source}}$}
                    [.{$\rho_{\,\text{l}/\edges}(\edges)$} ]
                    [.{$\rho_{\,\text{m}/\edges}(\edges)$} ]]
                [.{$\rho_{\,\text{r}/\edges}(\edges)$} ]]]]

  \end{subfigure}

  \caption{An example SQL query searching for paths of length three and its corresponding initial query plan.}
  \label{fig:join-plans}
\end{figure}

\begin{figure}[h]
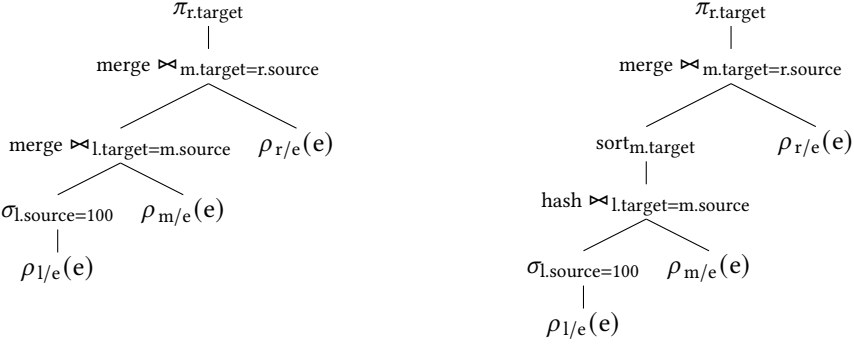

  \begin{subfigure}[t]{0.5\linewidth}
    \Tree[.{$\pi_{\text{r.target}}$}
        [.{$\text{\smaller merge} \bowtie_{\,\text{m.target}=\text{r.source}}$}
            [.{$\text{\smaller merge} \bowtie_{\,\text{l.target}=\text{m.source}}$}
                [.{$\sigma_{\text{l.source} = 100}$} [.{$\rho_{\,\text{l}/\edges}(\edges)$} ]]
                [.{$\rho_{\,\text{m}/\edges}(\edges)$} ]]
            [.{$\rho_{\,\text{r}/\edges}(\edges)$} ]]]
  \end{subfigure}%
  \begin{subfigure}[t]{0.5\linewidth}
    \Tree[.{$\pi_{\text{r.target}}$}
    [.{$\text{\smaller merge} \bowtie_{\,\text{m.target}=\text{r.source}}$}
    [.$\text{\smaller sort}_{\text{m.target}}$
    [.{$\text{\smaller hash} \bowtie_{\,\text{l.target}=\text{m.source}}$}
          [.{$\sigma_{\text{l.source} = 100}$} [.{$\rho_{\,\text{l}/\edges}(\edges)$} ]]
          [.{$\rho_{\,\text{m}/\edges}(\edges)$} ]]]
    [.{$\rho_{\,\text{r}/\edges}(\edges)$} ]]]
  \end{subfigure}%

  \caption{Optimized versions of the default query plan in \cref{fig:join-plans}. In the left plan, the selection on $\text{l.source} = 100$ has been pushed down past the joins. In the right plan, the bottom merge join on l and m has been replaced with a hash join; a sort has been added above to enforce that the inputs to the top merge join are sorted.}
  \label{fig:join-plans-opt}
\end{figure}

Database engines work by translating user-provided queries (often in SQL) down
to a query plan, then repeatedly applying rewrite rules to find more efficient
plans. For example, consider the SQL query in \cref{fig:join-plans}, which finds all nodes reachable in a path of exactly three steps from a source node with \ID{} $100$. The initial plan uses merge joins, which is reasonable if we assume that the relations are sorted or indexed on the join columns. But there is room for improvement, since the selection $\sigma_{\,\text{l.source} = 100}$ can be pushed down past the joins. The left side of \cref{fig:join-plans-opt} shows a plan where $\sigma_{\,\text{l.source} = 100}$ has been pushed down as
far as possible. This optimization can be discovered through equality
saturation with a rewrite rule that appropriately commutes selections with
joins.

However, such a plan may \emph{still} not be optimal; after selection, the
inputs to the bottom join may be relatively small. If relations are small and
fit in memory, then it may be more efficient to hash join them. Thus, we would
like to replace the merge join $\text{\small merge}
  \bowtie_{\,\text{l.target}=\text{m.source}}$ with a hash join $\text{\small
    hash} \bowtie_{\,\text{l.target}=\text{m.source}}$. But unconditionally
rewriting merge joins to hash joins is not valid---merge joins preserve sort
order, but hash joins do not. For example, if a merge join $J$ is being fed to
an operator which requires that its inputs be sorted (e.g. another merge join,
as in the left plan in \cref{fig:join-plans-opt}, or a non-commutative
aggregation), it is not valid to replace $J$ with a hash join.

It is not difficult to fix this though: we can add a sort \textit{enforcer}
\cite{volcano} underneath the top merge join that sorts its left input on the
$\text{m.target}$ column. Underneath the sort enforcer, it is
\textit{contextually valid} to rewrite merge joins to hash joins, because all
sub-plans will be eventually sorted on the $\text{m.target}$ column. The right
hand query plan in \cref{fig:join-plans-opt} shows the query plan with the sort enforcer added and bottom merge join
replaced with a hash join.

More generally, in database query optimization, query (sub-)plans must often
satisfy certain \textit{physical properties}, because their outputs are fed to
consumers who rely on those physical properties \cite{volcano}. Physical
properties include sort order, partitioning, and data location. In general,
valid rewrites must preserve physical properties. However, in certain contexts,
like underneath a sort enforcer, additional rewrites that ``break'' physical
properties become valid, because those physical properties will be
re-established by the enforcer.

\subsection{Simplifying conditionals}
\label{sec:conditionals}

As shown in the introduction, contextual equality saturation can also be used
to discover rewrites for conditional statements. The ternary $x == 2\;?\;(x
  \times y) : y$ can be rewritten to the ternary $x == 2\;?\;(y \gg 1) : y$,
since under the if branch of the ternary, we know that $x == 2$, which yields
the chain of equalities $x \times y \equiv 2 \times y \equiv y \gg 2$.

But not only can contextual equality saturation reason under conditionals, in
some circumstances, it can remove conditional statements entirely. Consider
another ternary $(a > b)\;?\;(a > b) : (a \leq b)$. With contextual
equivalences, one might reason:

\begin{enumerate}
  \item Under the then branch of the ternary, we have the additional equivalence $a > b
          \equiv \text{true}$. Hence, the e-class inside the then branch contains the
        terms $\{a > b, \text{true}\}$.
  \item Under the then branch of the ternary, we have the additional equivalence
        $\neg{}(a > b) \equiv \text{true}$. Another (general) rewrite rule could
        rewrite $\neg{} (a > b) \equiv b \leq a$, so, by transitivity, $b \leq a \equiv
          \text{true}$. The e-class inside the else branch would contain the terms
        $\{\neg (a > b), \text{true}, b \leq a\}$.
  \item An ``intersecting'' rule says if the bodies of \text{both} branches is
        equivalent to some term $t$, then the entire ternary is equivalent to $t$. That
        is, because \textit{both the e-classes} for the then branch and the else branch
        contain the term \text{true}, then the entire ternary is equivalent to
        \text{true}: $\big((a > b)\;?\;(a > b) : (a \leq b)\big) \equiv \text{true}$.
        (This rule is corresponds to the ``disjunction elimination'' or ``proof by
        cases'' inference rule in propositional logic \cite{colored-egraphs}.)
\end{enumerate}

We note that step 3 ``converts'' contextual equivalences within the branches of
the ternary into context-insensitive equivalences by taking the
``intersection'' of two equivalence relations. In order to support such
reasoning, a contextual equality saturation framework needs to let a user write
rules that can refer to and manipulate (contextual) equivalence relations.

\subsection{Lambda application}
\label{sec:lambda-app}

Finally, we show an example where contextual reasoning can simplify the body of
a lambda application. This type of contextual reasoning is similar to
conditional simplification, except has one additional complication. Consider
the lambda application $(\lambda{}x.x + 1)2$. We want to show that the entire
term is equal to $3$. Because $2$ is applied to $\lambda{}x.x + 1$, we can
contextually equate $x \equiv 2$ within the body of the lambda. Then, using a
context-insensitive rewrite rule $2 + 1 \equiv 3$, we can transitively find the
equivalence $x + 1 \equiv 3$, inside the lambda body. This means that the
entire lambda application $(\lambda{}x.x + 1)2 \equiv (\lambda x. 3)2$.

Here is the complication: we want to simplify the entire lambda application to
$3$, but so far we have only simplified the body. Just like in conditional
rewriting, how might we ``convert'' contextual equivalences on the body to
context-insensitive equivalences on the entire application? One might suggest
the following (incorrect) rule: if the lambda body is equal to a term $t$, then
the entire lambda application is equivalent to the term $t$. This is not
correct because the e-class of the body contains the term $t = x + 1$, and we
\textit{cannot} equate the whole lambda application to $x + 1$, since $x$ is
``unmentionable'' outside of the scope of the lambda.

The rule that we want is: if the lambda body is equal to a term $t$, and that
term $t$ does not transitively mention $x$, then the entire lambda application
is equal to $t$. Concretely, the e-class of the lambda body would contain the
terms $\{x+1, 2+1, 3\}$ when saturated. Under the above rule, we would
(correctly) equate the entire lambda application to the terms in $\{2+1, 3\}$.

However, this type of rule is hard to implement in existing equality saturation
frameworks because existing equality saturations frameworks like \egg{} and
\egglog{} \textit{canonicalize} the e-graph---the body of the lambda
application is not a set of terms, but instead an e-class. Furthermore, there
is no way to ``cleave apart'' an e-class to recover the terms that do not
contain $x$. Further research is needed to discover how contextual equality
saturation might support such an operation.\footnote{We note that existing
  equality saturations can implement beta-reduction (as well as entire
  interpreters for a lambda calculus) by using e-graph \textit{analyses} to keep
  track of free and bound variables in an expression.}

\section{A set-theoretic perspective}

In this section we present ongoing work on a set-theoretic perspective on
contextual equality saturation. We aim to describe contextual equality
saturation in an implementation-independent manner and to suggest how a
relational contextual equality saturation framework might practically support
contexts. We assume familiarity with lattices and equivalence relations;
definitions for lattices and equivalence relations can be found in the
\nameref{appendix}. We begin with a definition of \textit{quotient sets}, and
we relate them to e-classes and a hierarchy of equivalence relations.

\begin{definition}
  Let $A$ be a set, and let $\sim$ be an equivalence relation on $A$. The
  \textit{quotient set} $A/\sim$ is a set of equivalence classes where two
  elements $a, b \in A$ are in the same equivalence class iff $a \sim b$.
\end{definition}

\begin{example}
  Let $\Sigma$ be a set of function symbols, and let $R = \{f(a, b, \ldots) \mid
    f, a, b \ldots \in \Sigma \}$ be a set of terms. For two terms $a$ and $b$, let
  $a \sim b$ iff $a$ and $b$ are equivalent terms. Then $R/\sim$ is exactly the
  set of e-classes of $R$ under $\sim$, and the quotient map $\pi: R \to R/\sim$
  sends a term to its e-class (it is exactly the \texttt{\small find} operation
  in \egglog{} \cite{egglog}).
\end{example}

We note that $R/\sim$ is still not the ``e-graph'' stored by \egglog{} because
terms within an equivalence class in $R/\sim$ still refer to terms in $R$; that
is, $R/\sim$ is not canonicalized, so it may be exponentially large. We define
the canonicalized database $(R/\sim)^*$ as $(R/\sim)^* = \{\{f(\pi(a), \pi(b),
  \ldots), \ldots\} \mid {f(a, b, \ldots), \ldots} \in R/\sim\}$; i.e. we replace
references to subterms with references to the subterms' equivalence classes.
This shrinks the size of each equivalence class because terms in $R/\sim$ that
used to reference equivalent but not identical subterms are merged in
$(R/\sim)^*$.

\begin{proposition}
  Let $B^+$ denote the transitive closure of $B$. The set of equivalence
  relations on $A$ forms a lattice, where $S \meet T \defeq S \cap T$ and $S\join
    T \defeq (S \cup T)^+$. Furthermore, $\sim_1 \;\leq\; \sim_2$ iff $\sim_1
    \;\subseteq\; \sim_2$.
\end{proposition}

Intuitively, the meet of two equivalence relations $S$ and $T$ is another
equivalence relation $U$ where $a \sim b$ if $a$ is equivalent to $b$ under
\textit{both $S$ and $T$.} The join of two equivalence relations $S$ and $T$ is
another equivalence relation $V$ where $a \sim_V b$ if $a$ is equivalent to $b$
by \textit{chaining together any equivalences of $S$ or $T$.} (Transitive
closure formally captures the notion of ``chaining together.'')

\begin{definition}
  \label[definition]{def:labeled-lattice}
  Let $A$ be a set, let $\sim^{A}$ be the set of all equivalence relations on
  $A$, and let $L$ be a lattice with an element called bottom, $\bot \in L$,
  satisfying $\forall l \in L, \bot \leq l$. A \textit{context-annotated
    equivalence relation} is a mapping $\phi: L \to{} \sim^{A}$ where $\phi$
  preseves order: $l_1 \leq l_2 \implies \phi(l_1) \leq \phi(l_2)$.
\end{definition}

\Cref{def:labeled-lattice} suggests what structure contexts should have to be suitable for labels in
contextual equality saturation---they should have the property that as one
learns more information, more, not fewer, equalities become available. We
associate with the base context, labeled $\bot$, the fewest equivalences
$\phi(\bot)$---these are the rewrites that are allowed in every context. As one
traverses up the context lattice, more information is learned, and more terms
are equal.

In \cref{sec:query-opt}, our context lattice contained two elements: $\bot$
represented a context where sort order had to be preserved, so merge join and
hash join were not equivalent. Underneath a sort operator, however, we
transition up the context lattice to a context $s$ where we no longer had to
preserve sort order. Thus, we verify that $\phi(\bot) \leq \phi(s)$ (in fact,
the equality is strict: $\phi(\bot) < \phi(s)$).

Similarly, we (secretly) used the lattice structure of equivalence relations in
\cref{sec:conditionals}. Under each branch of the conditional, we transitioned
up the lattice: in the then branch, we transitioned from $\bot$ to $t > \bot$,
where $a > b \equiv \text{true}$. In the false branch, we transitioned up to $f
  > \bot$, where $\neg(a > b) \equiv \text{true}$. When we took the intersection
of the terms for both branches of the ternary, we applied meet ($\meet$) on the
contextual equivalence relations for each branch, ``updating'' $\phi(\bot)
  \leftarrow \psi(\bot)$ where $\psi(\bot) = \phi(\bot) \join (\phi(t) \meet \phi(f))$.

\vspace{0.3cm}

\noindent
\begin{minipage}{.72\textwidth}
  \setlength{\parindent}{1pc}
  \begin{proposition}
    \label[proposition]{prop:coarser}
    Let $A$ be a set, and let $\sim_1$ and $\sim_2$ be two equivalence relations on
    $A$ where $\sim_1 \;\leq\; \sim_2$. Then we have (note $1 \iff 2 \implies 3$):
    \begin{enumerate}
      \item The quotient map (which sends elements to their equivalence classes) $\pi_2 : R
              \to R/\sim_2$ factors through the quotient map $\pi_1 : R \to R/\sim_1$: there
            is a map $q: R/\sim_1 \to R/\sim_2$ such that $\pi_2 = q \circ \pi_1$ See
            \cref{fig:factors}.
      \item The quotient set $A/\sim_2$ is ``more coarse'' than the quotient set $A/\sim_1$
            in the following sense: every equivalence class $c \in A/\sim_2$ is a union of
            one or more equivalence classes $C \subseteq A/\sim_1$; that is, $c =
              \bigcup{C}$.
      \item $A/\sim_2$ has fewer equivalence classes: $\abs{A/\sim_2} \leq
              \abs{A/\sim_1}$.
    \end{enumerate}
  \end{proposition}
\end{minipage}
\hspace{0.015\textwidth}
\begin{minipage}{.23\textwidth}
  \vspace{-0.25cm}
  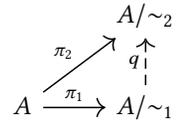
\begin{figure}[H]
    \centering
    \begin{tikzcd}
      & A/\sim_2                        \\
      A \arrow[r, "\pi_1"] \arrow[ru, "\pi_2"] & A/\sim_1 \arrow[u, "q", dashed]
    \end{tikzcd}

    \vspace{-0.3cm}

    \caption{$\pi_2$ factors through $\pi_1$ and $q$.}
    \label{fig:factors}
  \end{figure}
\end{minipage}

\vspace{0.3cm}

In \cref{prop:coarser}, one can interpret $\sim_1$ as some (possibly contextual)
equivalence relation and $\sim_2$ as a coarser equivalence relation. Item (1)
says that any database that stores a finer database of terms $R/\sim_1$ can
always ``recover'' the coarser set of terms by applying $q$, which sends
$R/\sim_1 \to R/\sim_2$.

As observed in both \egglog{} \cite{egglog} and colored e-graphs
\cite{colored-egraphs}, it can be exponentially faster to e-match on a
canonicalized e-graph versus an uncanonicalized e-graph. \emph{The existence of
  $q$ suggests that contextual equality saturation frameworks can trade off space
  for time}: instead of storing a canonicalized copy of the e-graph for every
equivalence relation, databases can store e-graphs for a \emph{lower-bound} of
contexts. When e-matching is requested on some context that is not explicitly
materialized, frameworks have two options: either they can e-match directly on
the stored e-graphs, incurring the cost of a join on the equivalence relation;
or they can apply $q$ on-the-fly and e-match on a canonicalized e-graph. If
$\sim_2$ adds many equivalences, then the framework will have to join on many
terms, so it is likely cheaper to apply $q$ up front to canonicalize (a copy
of) the e-graph to the requested equivalence. On the other hand, $\sim_2$ adds
very few equivalences, then a join might be cheaper. This type of reasoning is
the bread and butter of classical database management systems, which can
collect statistics to estimate the size of intermediate relations and
aggregations \cite{datacube}.

Finally, we note that item (3) in \cref{prop:coarser} implies that as more equivalences are added, the number of equivalence classes
in the canonicalized database can shrink. Relational logic systems like Datalog
rely on monotonicity to be efficient and correct. Thus, a relational equality
saturation framework that supports contexts needs to prevent users from
observing any non-monotonicity. For example, programs should not be able to
depend on how many equivalence classes there are in the database. (\egglog{}
already achieves this for a single equivalence relation.)

\section{Conclusion}

We have provided an overview of existing approaches to contextual equality
saturation, and shown three problems that contextual equality saturation may be
able to (partially) solve. Our early work points towards a set-theoretic model
for contextual equality saturation that lends itself to natural system
optimizations from the database literature.

In the future, we plan to further develop our set-theoretic model and extend it
to the relational setting. In addition, we aim to develop a syntax that lets
users easily express context-sensitive rewrites in a relational model and
manipulate equivalences across contexts, but prevents them from observing
non-monotonicity (which is crucial for performance and correctness). Finally,
we aim to explore whether existing Datalog systems like \egglog{} and
Souffl{\'e} can be adapted to support a hierarchy of equivalence relations.

\begin{acks}
  This work is supported in part by National Science Foundation CISE
  Expeditions Award CCF-1730628, IIS-1955488, IIS-2027575, GRFP Award
  DGE-2146752, DOE award DE-SC0016260, ARO award W911NF2110339, and ONR award
  N00014-21-1-2724, and by gifts from Astronomer, Google, IBM, Intel, Lacework,
  Microsoft, Mohamed Bin Zayed University of Artificial Intelligence, Nexla,
  Samsung SDS, Uber, and VMware.
\end{acks}

\bibliography{main}

\appendix

\section{Definitions}
\label{appendix}

\begin{definition}
  A \textit{meet semilattice} is an algebraic structure $(S, \meet)$ where
  $\meet$ (pronounced meet, or greatest lower bound) is a binary operator that is
  idempotent, commutative, and associative. Given a join semilattice, we can
  define $a \leq b$ if $a \meet b = a$ \cite{gratzer}.
\end{definition}

\begin{definition}
  Dually, a \textit{join semilattice} is an algebraic structure $(S, \join)$
  where $\join$ (pronounced join, or least upper bound) is a binary operator that
  is idempotent, commutative, and associative. Given a join semilattice, we can
  define $a \geq b$ if $a \join b = a$ \cite{gratzer}.
\end{definition}

\begin{definition}
  A \textit{lattice} is an algebraic structure $(S, \meet, \join)$ where $(S,
    \meet)$ is a meet semilattice, $(S, \join)$ is a join semilattice, and $\meet$
  and $\join$ satisfy the absorption law $a \meet (a \join b) = a \join (a \meet
    b) = a$.
\end{definition}

\begin{example}
  Let $S = \mathbb{Z}$, $\meet = \min$, and $\join = \max$. Then $(\mathbb{Z},
    \min, \max)$ is a lattice. For all $a, b \in \mathbb{Z}$:
  \begin{enumerate}
    \item Both $\min$ and $\max$ are idempotent, associative, and commutative.
    \item The meet of $a$ and $b$ is $\min(a, b)$, and if $\min(a, b) = a$ then $a \leq
            b$ (under the normal ordering).
    \item The join of $a$ and $b$ is $\max(a, b)$, and if $\max(a, b) = a$ then $a \geq
            b$.
    \item $\min(a, \max(a, b)) = a = \max(a, \min(a, b))$.
  \end{enumerate}
\end{example}

\begin{definition}
  A \textit{binary relation} on a set $A$ is a subset $R$ of $A \times A$. For
  $a, b \in A$, we write $a \sim{} b$ if $(a, b) \in R$. A binary relation is an
  \textit{equivalence relation} if it is reflexive, symmetric, and transitive
  \cite{dummit-foote}.
\end{definition}

\end{document}